\documentclass[10pt,conference]{IEEEtran}
\IEEEoverridecommandlockouts
\def\BibTeX{{\rm B\kern-.05em{\sc i\kern-.025em b}\kern-.08em
    T\kern-.1667em\lower.7ex\hbox{E}\kern-.125emX}}

\usepackage{hyperref}
\usepackage{adjustbox}
\usepackage{algorithm}
\usepackage{algorithmicx}
\usepackage{algpseudocode}
\usepackage{amsmath,amsfonts}
\usepackage{amssymb}
\usepackage{array}
\usepackage{balance}
\usepackage{booktabs}
\usepackage[skip=1pt,labelfont=bf]{caption}
\usepackage{calc}
\usepackage{calligra}
\usepackage{color}
\usepackage{colortbl}
\usepackage{courier}
\usepackage{csvsimple}
\usepackage{enumitem}
\usepackage{fancybox}
\usepackage{fontenc}
\usepackage{fontawesome5}
\usepackage{graphicx}
\usepackage{listings}
\usepackage{longtable}
\usepackage{lscape}
\usepackage{makecell}
\usepackage{marvosym}
\usepackage{moreverb}
\usepackage{multicol}
\usepackage{multirow}
\usepackage{pifont}
\usepackage{pgfplots}
\usepackage{rotating}
\usepackage{setspace}
\usepackage{siunitx}
\usepackage{soul}
\usepackage{subcaption}
\usepackage{tablefootnote}
\usepackage[most]{tcolorbox}
\usepackage{threeparttable}
\usepackage{tikz}
\usepackage[normalem]{ulem}
\usepackage{url}
\usepackage{wasysym}
\usepackage{xspace}

\usepgfplotslibrary{statistics}
\usetikzlibrary{matrix, shapes.geometric, arrows}
\usetikzlibrary{shapes, arrows, positioning}

\algnewcommand\algorithmicforeach{\textbf{for each}}
\algdef{S}[FOR]{ForEach}[1]{\algorithmicforeach\ #1\ \algorithmicdo}

\newcolumntype{L}[1]{>{\raggedright\let\newline\\\arraybackslash\hspace{0pt}}m{#1}}
\newcolumntype{C}[1]{>{\centering\let\newline\\\arraybackslash\hspace{0pt}}m{#1}}
\newcolumntype{R}[1]{>{\raggedleft\let\newline\\\arraybackslash\hspace{0pt}}m{#1}}

\definecolor{codegreen}{rgb}{0,0.6,0}
\definecolor{codered}{rgb}{1,0,0}
\definecolor{codegray}{rgb}{0.5,0.5,0.5}
\definecolor{codepurple}{rgb}{0.58,0,0.82}
\definecolor{backcolour}{rgb}{0.95,0.95,0.92}
\definecolor{lightgray}{gray}{0.9}

\newboolean{showcomments}
\setboolean{showcomments}{true}
\ifthenelse{\boolean{showcomments}}
 { \newcommand{\mynote}[2]{
      \fbox{\bfseries\sffamily\scriptsize#1}
        {\small$\blacktriangleright$\textsf{\emph{#2}}$\blacktriangleleft$}}}
        { \newcommand{\mynote}[2]{}}
        
\definecolor{DarkOrange}{rgb}{0.8,0.3,0.0}
\definecolor{DarkCyan}{rgb}{0.0, 0.55, 0.55}
\definecolor{DarkCyel}{rgb}{1.0, 0.49, 0.0}
\definecolor{yellow-green}{rgb}{0.6, 0.8, 0.2}

\newcolumntype{?}{!{\vrule width 1pt}}

\newcommand{\tool}{\textsc{InCTRL}\xspace}

\definecolor{lightgreen}{rgb}{0.56, 0.93, 0.56}
\newcommand{\find}[1]{
\begin{tcolorbox}[leftrule=0.4mm,rightrule=0mm,toprule=0mm,bottomrule=0mm,left=0.0pt,right=0.0pt,top=0pt,bottom=0pt]
\em #1
\end{tcolorbox}
}

\lstdefinelanguage{mymarkdown}{
    morekeywords={*,\#, \#\#, \#\#\#},
    sensitive=false,
    morecomment=[l]{//},
    morestring=[b]",
    commentstyle=\color{codegreen},
    keywordstyle=\color{magenta},
    numberstyle=\tiny\color{codegray},
    stringstyle=\color{codepurple},
    basicstyle=\small,
    breakatwhitespace=false,         
    breaklines=true,
    breakindent=0pt,
    keepspaces=true,                 
    numbers=left,                    
    numbersep=5pt,                  
    showspaces=false,                
    showstringspaces=false,
    showtabs=false,                  
    tabsize=2,
}

\lstdefinestyle{mystyle}{
    commentstyle=\color{codegreen},
    keywordstyle=\color{magenta},
    numberstyle=\small\color{black},
    stringstyle=\color{codepurple},
    basicstyle=\scriptsize\ttfamily,
    breakatwhitespace=false,
    breaklines=true,
    captionpos=b,
    keepspaces=true,
    showspaces=false,
    showstringspaces=false,
    showtabs=false,
    tabsize=2
}

\lstdefinelanguage{diff}{
  morecomment=[f][\color{blue}]{@@},     
  morecomment=[f][\color{red}]-,         
  morecomment=[f][\color{codegreen}]+,       
  morecomment=[f][\color{red}]{---}, 
  morecomment=[f][\color{codegreen}]{+++},
  numberstyle=\tiny\color{codegray},
  numbers=left,                    
  numbersep=5pt,         
}

\setlist{noitemsep} 

\definecolor{darkpastelred}{rgb}{0.76, 0.23, 0.13}
\definecolor{ao(english)}{rgb}{0.0, 0.5, 0.0}

\definecolor{darkpastelred}{rgb}{0.76, 0.23, 0.13}
\definecolor{ao(english)}{rgb}{0.0, 0.5, 0.0}

\newboolean{useblue}
\setboolean{useblue}{false} 

\newcommand{\maybeblue}[1]{%
    \ifthenelse{\boolean{useblue}}%
    {\textcolor{blue}{#1}}%
    {#1}%
}

\usepackage{pifont}
\usepackage{wrapfig}
\begin{document}

\title{In-Context Code-Text Learning for \\Bimodal Software Engineering}

\author{\IEEEauthorblockN{1\textsuperscript{st} Xunzhu Tang}
\IEEEauthorblockA{
\textit{University of Luxembourg}\\
Luxembourg, Luxembourg \\
xunzhu.tang@uni.lu}
\and
\IEEEauthorblockN{2\textsuperscript{nd} Liran Wang}
\IEEEauthorblockA{\textit{School of Computer Science and Engineering} \\
\textit{Beihang University}\\
Beijing, China \\
wanglr@buaa.edu.cn}
\and
\IEEEauthorblockN{3\textsuperscript{rd} Yonghui Liu}
\IEEEauthorblockA{\textit{Faculty of Information Technology} \\
\textit{Monash University}\\
Melbourne, Australia \\
yonghui.liu@monash.edu}
\and
\IEEEauthorblockN{4\textsuperscript{th} Linzheng Chai}
\IEEEauthorblockA{\textit{School of Computer}\\ \textit{Science and Engineering} \\
\textit{Beihang University}\\
Beijing, China \\
challenging@buaa.edu.cn}
\and
\IEEEauthorblockN{5\textsuperscript{th} Jian Yang}
\IEEEauthorblockA{\textit{School of Computer}\\ \textit{Science and Engineering} \\
\textit{Beihang University}\\
Beijing, China \\
jiaya@buaa.edu.cn}
\and
\IEEEauthorblockN{6\textsuperscript{th} Zhoujun Li}
\IEEEauthorblockA{\textit{School of Computer}\\ \textit{Science and Engineering} \\
\textit{Beihang University}\\
Beijing, China \\
lizj@buaa.edu.cn}
\and
\IEEEauthorblockN{7\textsuperscript{th} Haoye Tian}
\IEEEauthorblockA{\textit{School of Computing and Information Systems} \\
\textit{University of Melbourne}\\
Melbourne, Australia \\
haoye.tian@unimelb.edu.au}
\and
\IEEEauthorblockN{8\textsuperscript{th} Jacques Klein}
\IEEEauthorblockA{\textit{SnT Centre for Security, Reliability and Trust} \\
\textit{University of Luxembourg}\\
Luxembourg, Luxembourg \\
jacques.klein@uni.lu}
\and
\IEEEauthorblockN{9\textsuperscript{th} Tegawendé F. Bissyandé}
\IEEEauthorblockA{\textit{SnT Centre for Security, Reliability and Trust} \\
\textit{University of Luxembourg}\\
Luxembourg, Luxembourg \\
tegawende.bissyande@uni.lu}
}


\maketitle

\begin{abstract}
Bimodal software analysis initially appeared to be within reach with the advent of large language models. Unfortunately, the complex interplay of natural language text and code in software engineering, presents unique challenges that prevent pretrained models to generalize to a variety of tasks. We postulate that in-context learning for the code-text bimodality is a promising avenue. This paper thus introduces a comprehensive study of in-context code-text learning, focusing on leveraging pretrained CodeLLAMA models.

We consider a diverse dataset encompassing 23 software engineering tasks, which we transform in an in-context learning format. To effectively extract informative features, we propose a configurable prompt template. Our proposed pipeline, \tool, then unifies prompt learning across various software engineering tasks. Extensive evaluation on the study datasets demonstrates the superiority of \tool-models in few-shot performance, surpassing state-of-the-art models including the support model, CodeLLAMA. Typically, we observe that applied to the CodeLLAMA model, \tool brings improvements in terms of precision (at least about 12\%) and recall (up to 93.88\%) on various tasks. For example, on the task of program repair, \tool improves the BLEU score of CodeLLAMA by 85 points, while for clone detection, \tool achieves an improvement of 69 percentage points. Moreover, \tool-models offer state-of-the-art performance when using retrieval-augmented generation on individual downstream tasks. Finally, we qualitatively analyze the benefits of \tool over CodeLLAMA and open-source all models for broader impact. 

We make our code and dataset publicly available at:
\begin{center}
        {\url{https://anonymous.4open.science/r/inctrl-B65B}}
\end{center}

\end{abstract}

\section{Introduction}
\label{sec:introduction}

Software development is a complex endeavor characterized by its inherent interplay of code and natural language. While significant advancements have been made in both code analysis and natural language processing, effectively integrating these two modalities remains a critical challenge~\cite{casalnuovo2020theory}. Yet, this integration is essential for addressing complex tasks that require a deep understanding of both code syntax and semantic context. In this setting, Large Language Models (LLMs) appear promising due to their ability to process and generate both text and code, which offers unprecedented potential to automate tasks, improve code quality, and enhance developer productivity~\cite{hou2023large}.

A critical factor in LLM effectiveness is the quality and diversity of training data~\cite{wei2023magicoder, yu2024codereval}. While source code is the foundation, incorporating additional data types such as bug reports~\cite{kang2023large, kang2023large}, patches~\cite{tian2023best, zhang2024appt}, and vulnerable code snippets~\cite{tang2023just, zhou2024large} has proven invaluable for specific tasks such as program repair or vulnerability detection. Yet, building general-purpose code-text models remains challenging due to the varied nature of code and the complexities of bridging the gap between natural language and formal logic~\cite{hadi2023survey}. Indeed, source code is inherently bimodal, composed of both a formal algorithmic component and an informal natural language component consisting for example of identifiers, comments and other text artefacts~\cite{allamanis2018survey}. While research has predominantly examined these channels in isolation, attempts for comprehensive approaches using machine learning have been proposed in the literature~\cite{dash2018refinym}. The two channels are indeed interconnected, with natural language often providing context, explanations, or summaries of the underlying code. By jointly analyzing both channels, researchers can potentially enhance their understanding and analysis of software systems~\cite{allamanis2015bimodal}.
Unfortunately, building general-purpose code-text models presents unique challenges due to the diverse input distributions and task variations. Bimodal software engineering  further faces complex challenges for identifying synchronization points between code and text as well as managing noise, which come in the form of ambiguity in the natural language channel and imprecision in the code algorithmic channel. 

To effectively address the specificities of bimodal software engineering, LLMs often require carefully-crafted datasets. This process demands a deep understanding of downstream tasks, code-context comprehension, and code-text alignment to ensure consistency and quality across diverse task types. While essential for enhancing LLM performance in software engineering, current data generation methods, such as those outlined by  Luo \textit{et al}.~\cite{luo2023wizardcoder} and Wang \textit{et al}.~\cite{wang-etal-2023-self-instruct}, frequently rely on limited task sets or heuristic approaches. This reliance can inadvertently introduce systemic biases inherited from both the LLMs and the predefined tasks themselves. Indeed, previous approaches to addressing these challenges have primarily fallen into two categories: multitask learning~\cite{liu2020multi, li2022tale} and task-specific component extension~\cite{xia2023automated, zhang2023repocoder, wei2022clear}. Multitask learning attempts to formulate various code-based tasks into a uniform input-output format, offering versatility but often struggling to capture nuanced differences between diverse tasks. Task-specific component extension augments pre-trained LLMs with additional components trained on task-oriented data, allowing for more tailored solutions but facing challenges in scalability and generalization to novel tasks. Both approaches have demonstrated limitations in their ability to generalize effectively to unseen datasets and tasks.

{\bf This paper.} In response to the aforementioned challenges,  we propose a novel framework designed to significantly enhance LLM capabilities in handling diverse software engineering tasks. We refer to it as In-Context Code-Text Learning (\tool). Central to our approach is a unified code-text interface that generalizes task-oriented code-text instructions.  By consolidating code and text data into a single format, we significantly expand the training data available to LLMs, resulting in models that are more versatile and less biased when tackling software engineering problems.

Our research contributes significantly to the emerging field of in-context code-text learning, a critical component for achieving bimodal software analysis. By establishing a foundation for effective code-text interaction, we envision future systems capable of accurately understanding and generating both modalities in a contextually aware manner. Such models will be instrumental in bridging the human-computer divide, enabling fluent translation between high-level instructions and executable code.

Key aspects of our proposed methodology include:

\begin{itemize}
    \item \textbf{Unified Code-Text Interface:} We introduce a novel interface that bridges the gap between diverse code-related tasks, allowing for a more cohesive representation of various SE problems.
    \item \textbf{Task-Oriented Instruction Generalization:} Our approach focuses on generalizing task-specific instructions, enabling LLMs to better understand and adapt to a wider range of SE tasks.

\end{itemize}

Experimental results, based on seven typical software engineering tasks and applied on 23 datasets, provide sufficient data to support the claims that:

\begin{description}
    \item \ding{182} \tool enhances LLMs for software engineering tasks by effectively leveraging in-context learning without requiring model retraining. Our experiments show that \tool improves CodeLLAMA by 17.36\% to 85.3\% on code generation tasks, 5.74\% to 13.71 on Text generation tasks, 12.31\% to 95.66\% on Classification tasks.
 \item \ding{183} \tool significantly improves classification performance by incorporating a Retrieval-Augmented Generation  module, addressing a key limitation of traditional in-context learning. Our experiments show that the RAG
module contributed to an average increase of 1.8 points of the
F1-score and for some datasets can achieve up to 7.26 points of improvement.
 \item \ding{184}  \tool demonstrates strong performance across various software engineering tasks, particularly excelling in code generation and repair. Its impact, however, varies across tasks and datasets,
influenced by factors such as dataset complexity, input length
and the intrinsic nature of the tasks.
 \item \ding{185}  \tool’s structured prompt design provides rich in-context learning opportunities, as exemplified by its effectiveness in code summarization. 
\end{description}

\section{Background}
\label{sec:motivation}
A wide array of existing software engineering research typically specializes in one particular downstream task in either code analysis or natural language processing.

\begin{figure}[h]
    \centering
    \includegraphics[width=\linewidth]{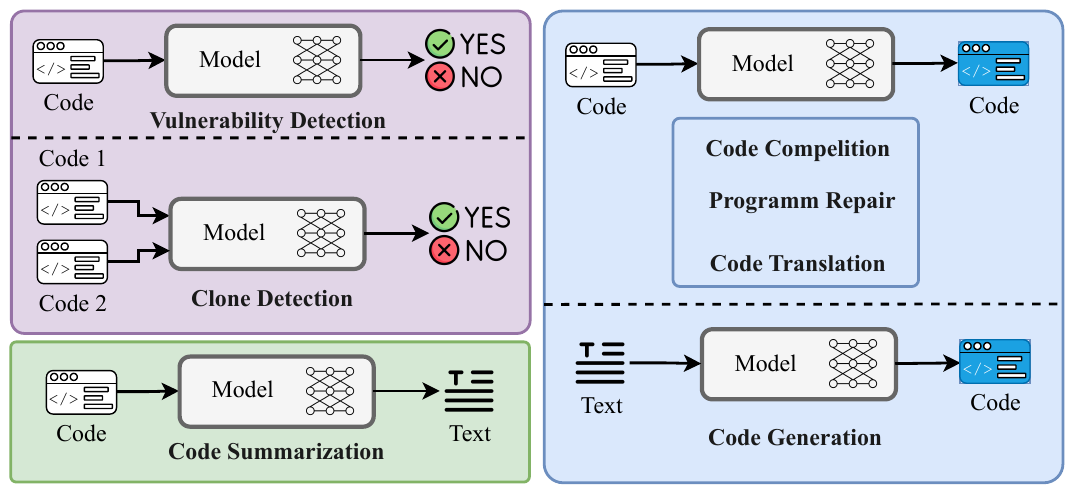}
    \caption{Illustration of models designed for SE tasks on code and text.}
    \label{fig:motimap}
\end{figure}

\textbf{Bimodal Software Engineering} represents an emerging field that seeks to bridge the gap between code and natural language within the software development lifecycle. This integration is essential for addressing complex tasks that require a deep understanding of both code syntax and semantic context. A primary challenge lies in aligning code and text representations effectively, as discrepancies can lead to errors, vulnerabilities, and hindered developer productivity~\cite{allamanis2018survey}. Traditional approaches often silo these modalities, as illustrated in Figure~\ref{fig:motimap}, limiting the potential synergies between them~\cite{casalnuovo2020theory}. Indeed, different tasks and datasets require various inputs and outputs, which require the creation and training of deep learning models that are capable of handling multiple downstream tasks, such as clone detection and code generation. The range of requirements underscores the need for more integrated and versatile tools in bimodal software engineering, capable of addressing the complex interplay between code and natural language across diverse software development scenarios.

To overcome these limitations, we propose a unified framework that leverages a general-purpose interface (i.e., an adapter), which helps to handle various software engineering tasks by seamlessly integrating code and text analysis. This approach empowers models to extract valuable insights from both domains, enhancing their capability to perform complex SE tasks like detecting vulnerabilities~\cite{Big-Vul, Draper}, explaining code\cite{CodeXGLUE, XLCoST}, and identifying code clones\cite{BigCloneBench} with greater accuracy and efficiency. Thus, our technique unifies prompt learning for large language models across various software engineering tasks, ensuring effective processing of diverse inputs to achieve state-of-the-art results.

\textbf{In-Context Learning} (ICL) is a key capability of large language models (LLMs) that allows them to adapt to new tasks without explicit training. By incorporating task demonstrations within the input prompt, LLMs can generate relevant outputs, as first highlighted with GPT-3~\cite{brown2020language}. Indeed GPT-3 demonstrated remarkable performance across a range of tasks by simply conditioning on task demonstrations embedded in the input context. Unlike traditional machine learning, ICL leverages the model's extensive pre-trained knowledge to perform tasks with minimal supervision, such as zero-shot or few-shot learning \cite{wei2022emergent}.  
Research has indicated that the effectiveness of ICL is influenced by factors such as the number, order, and quality of in-context examples~\cite{min2022rethinking, lu2021fantastically}. This sensitivity has prompted the development of prompt engineering techniques aimed at optimizing the presentation of information to the model \cite{reynolds2021prompt}. While the precise mechanisms underlying ICL are still under investigation, studies suggest that models may learn to identify patterns and correlations within the provided examples \cite{hancox2020robustness}.
Our work capitalizes on ICL's strengths, enabling rapid adaptation of the models by the \tool framework without extensive retraining.


\textbf{Multi-Task Learning} (MTL) is an advanced machine learning paradigm that contrasts with single-task learning approache. It involves the concurrent learning of multiple related tasks, leveraging shared information to enhance overall performance~\cite{zhang2021survey, li2022tale}. MTL offers a powerful framework for utilizing supervised data across related tasks, thereby reducing the reliance on task-specific labeled data~\cite{liu2022unified}. This approach inherently combats overfitting by encouraging the model to learn more generalizable features, effectively acting as a form of regularization~\cite{liu2020multi}. The application of MTL to boost task performance has been extensively explored and successfully implemented across diverse domains and scenarios.

While MTL enables models to learn multiple related tasks simultaneously, it offers rather a complementary perspective to our work. Indeed, our in-context code-text reasoning approach differs from MTL by emphasizing a flexible, task-agnostic interface capable of handling a wide range of software engineering problems.
 MTL indeed typically focuses on sharing parameters across predefined tasks to reduce overfitting and improve generalization.  While MTL typically involves predefined task structures with shared or task-specific parameters, the in-context code-text reasoning method emphasizes flexibility, allowing for the dynamic handling of various software engineering tasks through configurable prompt learning. We postulate that our method addresses more directly the unique challenges of bimodal software engineering, such as desynchronization between code and text, which are not explicitly targeted by traditional MTL techniques.

\textbf{Pre-trained Language Models} have significantly advanced natural language processing (NLP), consistently achieving state-of-the-art performance across a wide range of tasks~\cite{wang2021codet5, zhang2024appt, xia2023automated, fan2023large}. These models offer several key advantages: they extract universal representations from extensive corpora, provide enhanced model initialization for better generalization on downstream tasks, and serve as a form of regularization to combat overfitting when working with limited datasets. Researchers typically employ two main strategies to harness pre-trained language representations: feature-based approaches~\cite{von2022validity, wan2022they, wan2024deep}, which consider pre-trained representations as additional features, and fine-tuning approaches~\cite{weyssow2023exploring, huang2023empirical, zhang2024appt}, which adapt the entire pre-trained model for specific downstream tasks. However, these models often specialize in either \textit{classification} or \textit{generation} tasks. In contrast, we aim to take a more holistic view, unifying prompt learning across diverse software engineering tasks, and yielding models that are capable of addressing the multifaceted requirements of the software engineering field, including both code analysis and natural language processing aspects.

\section{\tool: In-Context Code-Text Learning}
\label{sec:approach}
The \tool framework is designed to integrate code and natural language text for various SE tasks, leveraging in-context learning to enhance model performance. This section details the construction of the \tool approach, its key components, and the configurations used.

\subsection{Architecture Overview}

\begin{figure}[ht]
    \centering
    \includegraphics[width=\linewidth]{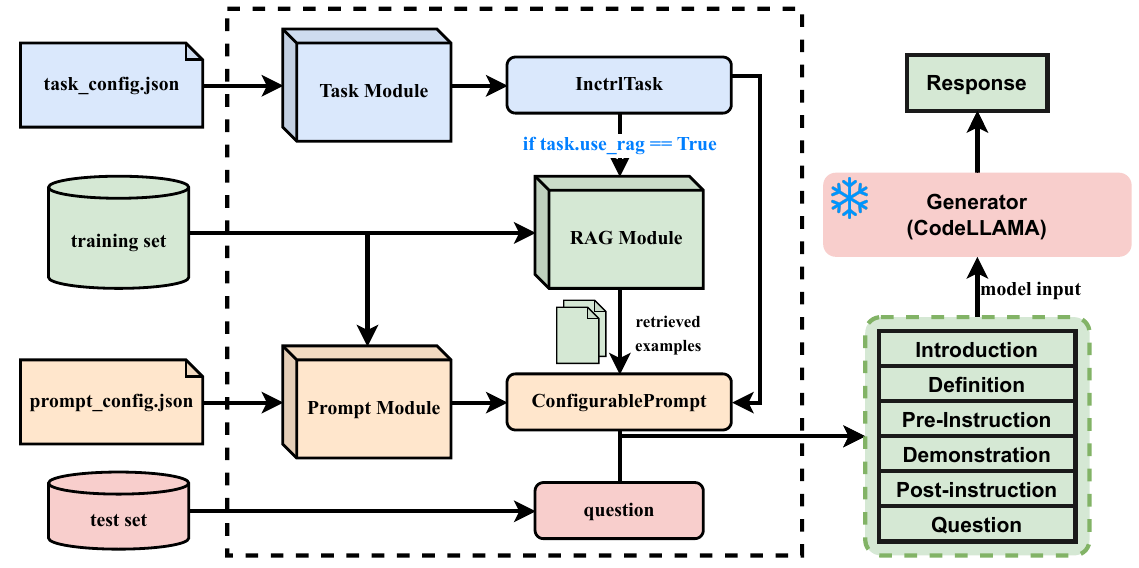}
    \caption{Architecture of \tool.}
    \label{fig:incctrl_architecture}
\end{figure}

The \tool architecture, depicted in Figure~\ref{fig:incctrl_architecture}, comprises several interconnected components designed to handle, together, complex software engineering tasks. 
A \textbf{Task Module} initializes task-specific settings based on the configuration file (\texttt{task\_config.json}). 
The \textbf{Prompt Module} loads prompt configurations (from \texttt{prompt\_config.json}), outlining the prompt structure and content for in-context learning. 
These configurations serve as the foundation for the \textbf{ConfigurablePrompt} component, which dynamically generates prompts by combining predefined templates with relevant examples retrieved by the Retrieval-Augmented Generation (\textbf{RAG}) \textbf{module}. This approach ensures that prompts are tailored to specific tasks, enhancing the model's ability to generate accurate and contextually appropriate responses.

If enabled, the {RAG Module} retrieves relevant examples from the training set. The retrieval process can be represented by the equation 
\begin{equation}
    \mathbf{E} = {RAG}(\mathbf{Q}, \mathbf{D})
\end{equation}
where $\mathbf{Q}$ is the query (task prompt) and $\mathbf{D}$ is the document corpus (training set). The retrieved examples $\mathbf{E}$ enrich the prompt. The \textbf{Question Module} handles specific queries posed by the user, integrating them into the configurable prompt to ensure comprehensive and tailored prompts for the task requirements. The enriched prompt is fed into the \textbf{Generator}, which, in our study, leverages the \textbf{CodeLLAMA} model to produce the final response. The model input, a combination of the prompt and the question, is represented as $\mathbf{R} = \text{Generator}(\mathbf{P} \oplus \mathbf{Q})$, where $\mathbf{R}$ is the generated response, $\mathbf{P}$ is the prompt, and $\mathbf{Q}$ is the question.


\subsection{Task Configuration}

\begin{figure}[!ht]
    \centering
    \includegraphics[width=\linewidth]{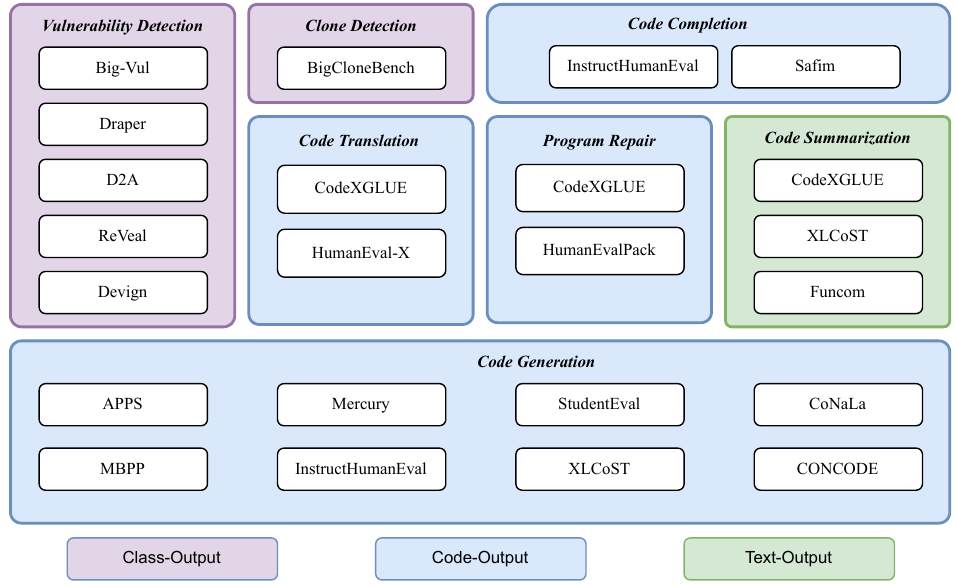}
    \caption{Software engineering tasks considered for our study}
    \label{fig:taskmap}
\vspace{-0.5cm}
\end{figure}

We consider three different types of tasks depending on the output: Class output, Code output, and text output. These tasks depicted in Figure~\ref{fig:taskmap} are executed across 23 datasets (cf. Section~\ref{exp_setup}.
The process begins with the {Task Module}, which initializes the specific task settings and determines whether to use the Retrieval-Augmented Generation (RAG) module based on the task requirements (specified in the configuration file).

The task configuration file contains parameters essential for defining the task environment. These include the task type (e.g., code generation, code summarization), input-output formats, evaluation metrics, and whether RAG is enabled. Let $\mathbf{T}$ denote the task settings vector derived from the configuration file, such that $\mathbf{T} = \{T_1, T_2, \ldots, T_n\}$, where $T_i$ represents individual configuration parameters. The Task Module parses $\mathbf{T}$ and initializes the environment accordingly.

If the task requires the use of the RAG module, the Task Module sets a binary flag $f_{RAG} \in \{0, 1\}$, where $f_{RAG} = 1$ indicates the activation of RAG. The decision rule can be formalized as follows:
\[ f_{RAG} = \begin{cases} 
1 & \text{if } T_i \text{ includes RAG-enabled tasks} \\
0 & \text{otherwise}
\end{cases} \]
This decision impacts subsequent steps, such as whether to retrieve examples from the training set.

The Task Module also initializes the input and output data structures. Let $\mathbf{I} = \{I_1, I_2, \ldots, I_m\}$ represent the input data configuration and $\mathbf{O} = \{O_1, O_2, \ldots, O_p\}$ the output data configuration. The mapping from input to output can be described by a function $\mathcal{F}$, which is parameterized by $\mathbf{T}$:
\[ \mathbf{O} = \mathcal{F}(\mathbf{I}; \mathbf{T}) \]
where $\mathcal{F}$ transforms the input data according to the task settings defined in $\mathbf{T}$.

Furthermore, the Task Module configures evaluation metrics to be used for assessing model performance. Let $\mathbf{M} = \{M_1, M_2, \ldots, M_q\}$ denote the set of metrics, with each $M_j$ representing a specific evaluation criterion, such as BLEU score for text generation or accuracy for classification tasks. The evaluation function $\mathcal{E}$ can be expressed as:
\[ \mathbf{E} = \mathcal{E}(\mathbf{O}; \mathbf{M}) \]
where $\mathbf{E}$ represents the evaluation results based on the generated output $\mathbf{O}$ and the selected metrics $\mathbf{M}$.

By systematically configuring the task environment, input-output mappings, and evaluation criteria, the Task Module ensures that the \tool framework is correctly initialized for the specific requirements of each software engineering task.

\subsection{Prompt Configuration}
Concurrently, the {Prompt Module} loads the prompt configurations from the configuration file. These configurations define the structure and content of the prompts used for in-context learning. The prompt configurations are crucial for generating task-specific prompts that guide the model in understanding and processing the input data effectively. Let $\mathbf{P}$ denote the prompt configuration vector derived from the configuration file, such that $\mathbf{P} = \{P_1, P_2, \ldots, P_k\}$, where $P_i$ represents individual configuration parameters, including template structures, placeholders for input variables, and instructions for the model.

The Prompt Module parses $\mathbf{P}$ to create a prompt template $\mathcal{T}$ dynamically adjusted based on the input data $\mathbf{I}$. Formally, let $\mathcal{T}$ be a function of $\mathbf{P}$ and $\mathbf{I}$: $\mathcal{T} = \mathcal{T}(\mathbf{P}, \mathbf{I})$, where $\mathcal{T}$ represents the structured prompt used to guide the model. To generate the final prompt $\mathbf{Pr}$, the Prompt Module substitutes the placeholders in $\mathcal{T}$ with actual values from $\mathbf{I}$: $\mathbf{Pr} = \mathcal{T}(\mathbf{P}, \mathbf{I})$. This ensures that the model receives a coherent and contextually relevant prompt.

Additionally, the prompt configuration includes rules for augmenting the prompts with supplementary context, such as examples or clarifying information. Let $\mathcal{A}$ denote the augmentation function: $\mathbf{Pr}' = \mathcal{A}(\mathbf{Pr}, \mathbf{C})$, where $\mathbf{Pr}'$ is the enriched prompt and $\mathbf{C}$ represents the context before question. 


\subsection{Configurable Prompt Generation}
\label{configurable_prompt}
The {ConfigurablePrompt} component is central to the \tool framework. It takes inputs from the Prompt Module and any additional context provided by the RAG module if enabled. The RAG module retrieves relevant examples from the training set, enriching the prompt with contextual information. The resulting prompt is a combination of predefined structures and dynamically retrieved examples, making it highly adaptable to various tasks.

Formally, let $\mathbf{Pr}$ represent the initial prompt generated by the Prompt Module, and $\mathcal{R}(\mathbf{I})$ denote the function of the RAG module that retrieves a set of relevant examples $\mathbf{E} = \{E_1, E_2, \ldots, E_n\}$ based on the input data $\mathbf{I}$. The \textit{ConfigurablePrompt} component combines these inputs to form an enriched prompt $\mathbf{Pr}'$. The combination process can be expressed as follows:
\begin{equation}
\mathbf{Pr}' = \mathcal{F}(\mathbf{Pr}, \mathbf{E}) = \mathbf{Pr} \cup \left( \bigcup_{i=1}^{n} E_i \right)
\end{equation}
where $\mathcal{F}$ is a function that integrates the initial prompt $\mathbf{Pr}$ with the retrieved examples $\mathbf{E}$, and $\cup$ denotes the union operation that appends the examples to the prompt.

Finally, when performing inference, we treat each a certain data entry in $\mathbf{Q_i}$ in the test set as a question, concatenate with $\mathbf{Pr}'$ to obtain the final prompt of this date entry $\mathbf{Pr}^{final}_{i}$ that is used as input to the model:

\begin{equation}
\mathbf{Pr}^{final}_{i} = \mathbf{Pr}' \oplus \mathbf{Q_i}.
\end{equation}

After conducting extensive experiments, we have summarized a high-quality template construction scheme and set it as the default structure for configurable prompt. Specifically, \tool's default scheme divides the configurable prompt into the following six sections (as previously illustrated in the lower right part of Figure~\ref{fig:incctrl_architecture}):
\begin{itemize}
    \item \textbf{INTRODUCTION}: provides a background context for the generator;
    \item \textbf{DEFINITION}: gives the basic requirements of the task;
    \item \textbf{PRE-INSTRUCTION}: gives more detailed requirements about the task;
    \item \textbf{DEMONSTRATION}: provides several input and output examples;
    \item \textbf{POST-INSTRUCTION}: emphasizes more detailed requirements about the task again;
    \item \textbf{QUESTION}: contains the query.
\end{itemize}

An example configurable prompt is provided in Figure~\ref{fig:casetext} for the task of code summarization. 


\subsection{Response Generation}

Once the final prompt is constructed, it is fed into the {Generator} module (i.e., CodeLLAMA in our case). The Generator leverages the structured prompt to produce accurate and contextually relevant responses. The response generation process can be modeled as a function $\mathcal{G}$ that maps the augmented prompt to a response vector, $\mathbf{R} = \mathcal{G}(\mathbf{Pr}^{final})$, ensuring that the output aligns with the user's query. In the context of code-related tasks, $\mathcal{G}$ utilizes the underlying capabilities of CodeLLAMA to perform various operations such as code generation, code completion, and providing natural language explanations. The generated response $\mathbf{R}$ can be further refined to ensure precision and relevance. For example, if the task involves generating a function definition based on a query, the response might include syntactic checks and semantic validations to ensure the code is both correct and optimal.

To enhance the robustness of the response, an additional post-processing step can be applied, incorporating feedback loops and iterative improvements. Let $\mathcal{P}$ be the post-processing function that refines the response, such that the finalized response is $\mathbf{R}^{\text{final}} = \mathcal{P}(\mathbf{R})$. This comprehensive approach ensures that the \tool framework can handle a wide range of tasks efficiently, producing high-quality outputs that meet the users' requirements in both code and natural language contexts.

\section{Inference}
The \tool framework is designed to accommodate a diverse range of software engineering tasks, including both classification and generation problems. Leveraging its modular architecture and effective prompt handling, \tool addresses these tasks through tailored inference strategies.

For generation-oriented tasks, such as code generation and open-ended code comprehension, the model is directly prompted to produce outputs. These generated sequences are then evaluated against ground truth using metrics like BLEU. In contrast, classification tasks, including vulnerability detection and clone detection, utilize a vocabulary ranking approach. The model generates potential candidates, and the option with the highest log-likelihood is selected as the final prediction. To enhance performance, especially for binary classification, we expand label sets to include semantically similar terms (e.g., "yes" and "true" for positive class), aligning the model's output with natural language patterns.

Mathematically, the classification process is expressed as:

\begin{equation}
\hat{y} = \arg\max_{c \in C} \log P(c | \mathbf{x})
\end{equation}
where $\hat{y}$ is the predicted class, $C$ is the set of candidate classes, and $\mathbf{x}$ is the input prompt.


For generation tasks, the generator produces a sequence of tokens, $\mathbf{y}$, given an input prompt, $\mathbf{x}$:

\begin{equation}
P(\mathbf{y} | \mathbf{x}) = \prod_{t=1}^{T} P(y_t | \mathbf{y}_{<t}, \mathbf{x})
\end{equation}
where $T$ is the sequence length, and $P(y_t | \mathbf{y}_{<t}, \mathbf{x})$ is the probability of generating token $y_t$ given previous tokens $\mathbf{y}_{<t}$  and the input prompt $\mathbf{x}$.

{\bf Prompt for baseline CodeLLAMA.} Experiments without \tool apply simple prompts that define the task and provide the input data. For example, for vulnerability detection, the prompt is as follows ``{\em Given the following code snippet, classify whether it is vulnerable or not. {\tt <code>} }''. All prompts used are in the replailcation package.

\section{Experimental Setup}
\label{exp_setup}

\subsection{Research Questions}
\begin{itemize}

    \item \textbf{RQ1: Effectiveness of in-context learning.} Considering the various code-text tasks in software engineering, to what extent model performance can be improved by in-context learning with \tool?

    \item \textbf{RQ2: Impact of Retrieval-Augmented Generation.} To what extent is the use of RAG for selecting examples for prompting necessary or beneficial?

    \item \textbf{RQ3: Performance on specific datasets.} From a fine-grained perspective, how does \tool influence the model's performance on individual datasets for each task?

    \item \textbf{RQ4: Qualitative case study.}  From the actual input-output examples of \tool, what qualitative assessments can we provide regarding its capabilities?
    
\end{itemize}

\subsection{Tasks and Datasets}
\label{datasets}
In Table~\ref{detailed_datasets}, we provide detailed information about the datasets used in our experimental setup. These datasets were collected from Huggingface. It is nonetheless worth noting that, for some datasets, we only used a subset. For those cases the  ``Remark'' column indicates which subset was considered.

\begin{table*}[ht]
\centering
\captionsetup{justification=centering}
\caption{Datasets considered for evaluating \tool on common software engineering tasks.}

\resizebox{.8\linewidth}{!}{
\begin{tabular}{l|lll}
\toprule
\textbf{Task} & \textbf{Dataset} & \textbf{Huggingface Repository ID} & \textbf{Remark} \\ \midrule
\multirow{5}{*}{Vulnerability Detection} 
& Big-Vul~\cite{Big-Vul} & benjis/bigvul &  \\
& Draper~\cite{Draper} & claudios/Draper &  \\
& D2A~\cite{D2A} & claudios/D2A & subset=``function'' \\
& ReVeal~\cite{ReVeal} & claudios/ReVeal &  \\
& Devign~\cite{Devign} & code\_x\_glue\_cc\_defect\_detection &  \\ \hline
Clone Detection & BigCloneBench~\cite{BigCloneBench} & code\_x\_glue\_cc\_clone\_detection\_big\_clone\_bench &  \\ \hline
\multirow{2}{*}{Code Completion} 
& InstructHumanEval~\cite{instructhumaneval} & codeparrot/instructhumaneval &  \\
& Safim~\cite{safim} & gonglinyuan/safim &  \\ \hline
\multirow{2}{*}{Code Translation} 
& CodeXGLUE~\cite{CodeXGLUE} & code\_x\_glue\_cc\_code\_to\_code\_trans &  \\
& HumanEval-X~\cite{HumanEval-X} & THUDM/humaneval-x & java-to-python \\ \hline
\multirow{2}{*}{Program Repair} 
& CodeXGLUE~\cite{CodeXGLUE} & ayeshgk/code\_x\_glue\_cc\_code\_refinement\_annotated &  \\
& HumanEvalPack~\cite{HumanEvalPack} & bigcode/humanevalpack & subset=``python'' \\ \hline
\multirow{3}{*}{Code Summarization} 
& CodeXGLUE~\cite{CodeXGLUE} & code\_x\_glue\_ct\_code\_to\_text & subset=``java'' \\
& XLCoST~\cite{XLCoST} & codeparrot/xlcost-text-to-code & subset=``C++-program-level'' \\
& Funcom~\cite{Funcom} & apcl/funcom-java-long &  \\ \hline
\multirow{8}{*}{Code Generation} 
& APPS~\cite{APPS} & codeparrot/apps & difficulties=``all'' \\
& MBPP~\cite{MBPP} & mbpp & subset=``sanitized'' \\
& Mercury~\cite{Mercury} & Elfsong/Mercury &  \\
& InstructHumanEval~\cite{instructhumaneval} & codeparrot/instructhumaneval &  \\
& StudentEval~\cite{StudentEval} & wellesley-easel/StudentEval &  \\
& XLCoST~\cite{XLCoST} & codeparrot/xlcost-text-to-code & subset=``C++-program-level'' \\
& CoNaLa~\cite{conala} & neulab/conala & subset=``curated'' \\
& CONCODE~\cite{CONCODE} & AhmedSSoliman/CodeXGLUE-CONCODE &  \\ \bottomrule
\end{tabular}}

\label{detailed_datasets}
\vspace{-0.5cm}
\end{table*}

We categorize most of the software engineering tasks into three types based on the modality of output data: \textit{code-output}, \textit{text-output}, and \textit{class-output}. The tasks that fall under the \textit{code-output} category include code generation, code translation, code completion, and program repair, encompassing 14 datasets (~\cite{APPS, MBPP, Mercury, instructhumaneval, StudentEval, XLCoST, conala, CONCODE, CodeXGLUE, HumanEval-X, HumanEvalPack}). The \textit{text-output} category consists of the code summarization task, which includes 3 datasets (~\cite{CodeXGLUE, XLCoST, Funcom}). The tasks in the \textit{class-output} category are vulnerability detection and clone detection, covering 6 datasets (~\cite{Big-Vul, Draper, D2A, ReVeal, Devign, BigCloneBench}). 


\subsection{Models}
\label{models}
We use CodeLLaMA as the base LLM. We select this LLM because \ding{182} it is open source and \ding{183} it is among the most popular in software engineering research. 
Specifically, we evaluated four sets of model parameters: CodeLLaMA-7B, CodeLLaMA-7B-Instruct, CodeLLaMA-13B, and CodeLLaMA-13B-Instruct. The difference between models with and without ``Instruct`` in their name lies in how they were trained and the type of data they were exposed to during the training process. Furthermore, we choose CodeBERT~\cite{codebert}, a pre-trained language model designed to understand and represent programming languages, as the embedding model in the RAG module. All the model parameters we used were downloaded from the official Hugging Face model repository.

\subsection{Metrics}
\label{metrics}
\subsubsection{Generation Tasks}
For \textit{code-output} and \textit{text-output} tasks, since they are both text generation tasks, we adopt \textbf{BLEU} \cite{bleu} as the evaluation metric. It is an automatic metric for evaluating the quality of machine-generated text by comparing it to reference, calculating a similarity score based on the overlap of n-grams between the generated text and the reference text. In the evaluation, we use the smoothed BLEU metric with 1-gram.

\subsubsection{Classification Tasks}
For \textit{class-output} tasks, we adopt \textbf{precision}, \textbf{recall}, and \textbf{F1-score} as evaluation metrics. All the \textit{class-output} tasks involved in our evaluation are binary classification tasks (i.e., outputting ``yes'' or ``no''). Under this setting, precision measures the accuracy of positive predictions, recall measures the completeness of positive predictions, and F1-score is the harmonic mean of precision and recall, providing a balanced measure of the classifier's performance.

\subsection{Implementation}
The RAG module of \tool is powered by {LlamaIndex}~\cite{llamaindex}. The evaluation of \tool is implemented based on {VLLM}~\cite{vllm} to ease reproducibility. For each dataset considered in this work, we provide its corresponding configurations to automatically build the interfaces. All experiments were conducted on a server equipped with 4 NVIDIA Tesla A800-80GB GPUs.
\section{Result and Analysis}
In this section, we present the overall in-context learning effectiveness of \tool in Sec.~\ref{rq1}, the necessity and effectiveness of the RAG module in Sec.~\ref{rq2}, the performance on individual datasets in Sec.~\ref{rq3} and the qualitative case study in Sec.~\ref{rq4}

\begin{table*}[!t]
\centering
\captionsetup{justification=centering}
\caption{Overall effectiveness of \tool across three types of tasks. }

\resizebox{0.7\linewidth}{!}{
\begin{tabular}{@{}l|cccc|c|cccccc@{}}
\toprule
                              & \multicolumn{4}{c|}{Code-Output} & Text-Output & \multicolumn{6}{c}{Class-Output}                \\ \midrule
\multirow{2}{*}{Model}        & CG     & CT     & CC     & PR    & CS              & \multicolumn{3}{c}{VD} & \multicolumn{3}{c}{CD} \\
                              & BLEU   & BLEU   & BLEU   & BLEU  & BLEU            & P     & R     & F1     & P      & R     & F1    \\ \midrule
codellama-7b                  & 3.36   & 1.53   & 3.12   & 1.47  & 0.16            & 0.0    & 0.0   & 0.0   & 0.0    & 0.0   & 0.0   \\
\multicolumn{1}{r|}{+ InCRTL} & \cellcolor{black!25}20.72  & \cellcolor{black!25}32.25  & \cellcolor{black!25}48.91  & \cellcolor{black!25}79.06 & \cellcolor{black!25}5.9             &  \cellcolor{black!25}12.31     &   \cellcolor{black!25}93.88    &    \cellcolor{black!25}21.77    & \cellcolor{black!25}14.11  & \cellcolor{black!25}95.66 & \cellcolor{black!25}24.59 \\ 

codellama-7b-Instruct               & 2.56   & 1.48   & 0.98   & 1.42  & 0.21            & 0.0    & 0.0   & 0.0   & 0.0    & 0.0   & 0.0   \\
\multicolumn{1}{r|}{+ InCTRL} & \cellcolor{black!25}26.94  & \cellcolor{black!25}59.85  & \cellcolor{black!25}46.97  & \cellcolor{black!25}86.72 & \cellcolor{black!25}13.92           &   \cellcolor{black!25}20.93    &   \cellcolor{black!25}74.79    &   \cellcolor{black!25}32.71     & \cellcolor{black!25}16.11  & \cellcolor{black!25}58.89 & \cellcolor{black!25}25.3  \\ 

codellama-13b                 & 2.2    & 1.65   & 0.7    & 1.41  & 0.5             & 0.0    & 0.0   & 0.0   & 0.0    & 0.0   & 0.0   \\
\multicolumn{1}{r|}{+ InCTRL} & \cellcolor{black!25}19.92  & \cellcolor{black!25}52.54  & \cellcolor{black!25}58.03  & \cellcolor{black!25}80.92 & \cellcolor{black!25}13.11           &   \cellcolor{black!25}22.5    &  \cellcolor{black!25}52.71     &    \cellcolor{black!25}31.62    & \cellcolor{black!25}16.6   & \cellcolor{black!25}66.61 & \cellcolor{black!25}26.58 \\ 

codellama-13b-Instruct              & 1.85   & 1.88   & 1.36   & 1.36  & 0.86            & 0.0    & 0.0   & 0.0   & 0.0    & 0.0   & 0.0   \\
\multicolumn{1}{r|}{+ InCTRL} & \cellcolor{black!25}32.18  &\cellcolor{black!25} 55.76  & \cellcolor{black!25}50.7   & \cellcolor{black!25}80.55 & \cellcolor{black!25}7.83            &  \cellcolor{black!25}13.3     &  \cellcolor{black!25}54.97     &   \cellcolor{black!25}21.48     & \cellcolor{black!25}17.64  & \cellcolor{black!25}69.66 & \cellcolor{black!25}28.15 \\ 
\bottomrule

\end{tabular}}

\label{table_overall}

\begin{tablenotes}
\footnotesize
    \item ``-Instruct'' indicates the instruct version of the model.
    \item abbreviations: code generation (CG), code translation (CT), code completion (CC), program repair (PR), code summarization (CS), vulnerability detection (VD) and clone detection (CD).
    \item P means precision; R means recall.
\end{tablenotes}
\vspace{-0.5cm}
\end{table*}

\subsection{RQ1: Effectiveness of in-context learning}
\label{rq1}

In this subsection, we investigate the overall performance of \tool on software engineering tasks. We first obtain results on all datasets mentioned in \ref{datasets} using models with and without applying \tool. Then, we aggregate the output results to compute the overall performance metrics for each task type. For models \tool, the prompt consists solely of a task definition and input data. 


Table~\ref{table_overall} presents the performance of models across three major categories of \textit{code-output}, \textit{text-output}, \textit{class-output}, subdivided into the seven specific tasks considered in our work: code generation (CG), code translation (CT), code completion (CC), program repair (PR), code summarization (CS), vulnerability detection (VD) and clone detection (CD). It can be observed that for all seven tasks, \tool can significantly enhance the capabilities of the models. Specifically, the smallest improvement is a 17-point increase in CG, while the largest is an 85-point surge in PR. This enhancement transforms models from nearly ineffective to highly effective, demonstrating \tool's impact on \textit{code-output} tasks.
Additionally, while \tool still marks substantial improvements over the baseline performance for \textit{text-output} tasks, it does not perform well on such tasks, compared with \textit{code-output} tasks. Finally, for \textit{class-output} tasks, direct model inference was initially completely unfeasible (no successful case) for CodeLLAMA on the studied datasets. \tool has therefore equipped the models with the capability to tackle such tasks.

From the table, we observe that without the enhancement provided by \tool, the instruct versions of the models  typically underperform compared to their non-instruct counterparts. This underperformance is attributed to the instruct versions requiring specific instructions, which are not directly provided by the initial prompts. However, with the application of \tool, the instruct versions outperform the non-instruct models in terms of CG, CT and PR. This improvement suggests that \tool effectively activates the instruct models' inherent capacity to learn and solve problems based on the configurable prompts provided by \tool. Specifically, \tool improves CodeLLAMA by 17.36\% to 85.3\% on Code-Output tasks, 5.74\% to 13.71 on Text-Output tasks, 12.31\% to 95.66\% on Class-Output tasks.

In RQ4, through an example, we will more intuitively analyze how the prompts generated by \tool activate the models.




\find{{\bf \ding{45} Answer to RQ1: }$\blacktriangleright$ 
\tool effectively leverages in-context learning to empower LLMs for handling complex bimodal software engineering tasks involving both code and text without the need for model retraining or fine-tuning.
$\blacktriangleleft$ }
\begin{table}[t]
\centering
\caption{Ablation study results: Performance (F1) with and without the RAG module.}
\resizebox{0.7\linewidth}{!}{

\begin{tabular}{@{}l|lllll@{}}
\toprule
Dataset                    & \tool +        & \multicolumn{1}{c}{7b}    & \multicolumn{1}{c}{7b-In} & \multicolumn{1}{c}{13b}   & \multicolumn{1}{c}{13b-In} \\ \midrule
\multirow{2}{*}{BCB~\cite{BigCloneBench}}    & w/o RAG & \multicolumn{1}{c}{23.72} & \multicolumn{1}{c}{22.58} & \multicolumn{1}{c}{23.21} & \multicolumn{1}{c}{21.12}  \\
                        & w/ RAG  & \multicolumn{1}{c}{24.59} & \multicolumn{1}{c}{25.3}  & \multicolumn{1}{c}{28.15} & \multicolumn{1}{c}{28.15}  \\ \midrule
\multirow{2}{*}{Devign~\cite{Devign}} & w/o RAG & \multicolumn{1}{c}{62.74} & \multicolumn{1}{c}{62.50} & \multicolumn{1}{c}{39.53} & \multicolumn{1}{c}{56.33} \\
                        & w/ RAG  & \multicolumn{1}{c}{62.84} & \multicolumn{1}{c}{62.80} & \multicolumn{1}{c}{25.77} & \multicolumn{1}{c}{39.83}  \\ \midrule
\multirow{2}{*}{D2A~\cite{D2A}}    & w/o RAG & \multicolumn{1}{c}{67.72} & \multicolumn{1}{c}{60.08} & \multicolumn{1}{c}{43.68} & \multicolumn{1}{c}{46.09} \\
                        & w/ RAG  & \multicolumn{1}{c}{68.14} & \multicolumn{1}{c}{68.23} & \multicolumn{1}{c}{55.46} & \multicolumn{1}{c}{54.78} \\ \midrule
\multirow{2}{*}{BigVul~\cite{Big-Vul}} & w/o RAG & \multicolumn{1}{c}{6.67} & \multicolumn{1}{c}{6.64} & \multicolumn{1}{c}{12.24} & \multicolumn{1}{c}{12.79} \\
                        & w/ RAG  & \multicolumn{1}{c}{7.11} & \multicolumn{1}{c}{11.56} & \multicolumn{1}{c}{17.36} & \multicolumn{1}{c}{16.55} \\ \midrule
\multirow{2}{*}{ReVeal~\cite{ReVeal}} & w/o RAG & \multicolumn{1}{c}{7.61} & \multicolumn{1}{c}{17.29} & \multicolumn{1}{c}{16.25} & \multicolumn{1}{c}{17.87} \\
                        & w/ RAG  & \multicolumn{1}{c}{12.05} & \multicolumn{1}{c}{19.73} & \multicolumn{1}{c}{18.56} & \multicolumn{1}{c}{19.92} \\ \midrule
\multirow{2}{*}{Draper~\cite{Draper}} & w/o RAG & \multicolumn{1}{c}{11.86} & \multicolumn{1}{c}{16.83} & \multicolumn{1}{c}{14.06} & \multicolumn{1}{c}{16.95} \\
                        & w/ RAG  & \multicolumn{1}{c}{11.87} & \multicolumn{1}{c}{17.12} & \multicolumn{1}{c}{17.69} & \multicolumn{1}{c}{17.08} \\ \bottomrule
\end{tabular}

}

\label{table_rag}

\begin{tablenotes}
\footnotesize
    \item ``-In'' indicates the instruct version of the model.
\end{tablenotes}
\vspace{-0.5cm}
\end{table}

\subsection{RQ2: Impact of Retrieval-Augmented Generation}
\label{rq2}

Unlike generation tasks (where randomly selecting samples from the training set to construct in-context prompts can enable the model to have a certain ability to complete the task), classification tasks have higher requirements for the relevance between the example (i.e., demonstration and the question. Directly using randomly sampled data as examples usually cannot make the model output valid results~\cite{llm4vd}. Therefore, we consider whether we can introduce the RAG module to make \tool also adaptable to \textit{class-output} tasks. Specifically, after enabling the RAG module, \tool treats the questions in the test set as queries and each data sample in the training set as a document to be retrieved. It then uses the cosine similarity of the semantic vectors (embeddings provided by CodeBERT) to retrieve the most relevant training samples to the question, which are used to fill the demonstration slots in the prompt. 

Table ~\ref{table_rag} shows the results of the ablation study for the RAG module of \tool on \textit{class-output} tasks. It should be noted that for some datasets with a large amount of training data, the time cost of retrieving from the full training set is too high. Therefore, we set a ratio $r$ to perform retrieval only from a subset of the training set. Specifically, if there are $n_{train}$ data points in the training set, we first randomly sample $r*n_{train}$ data instances and then perform RAG. For different datasets, the value of $r$ that achieves the best RAG effect varies. In Table~\ref{table_rag}, we have omitted the values of $r$ and default to the best results.

It can be seen that the RAG module of \tool almost always improves the model's performance on various classification tasks. 
For the classification tasks across six datasets, the RAG module demonstrated improved performance on five of them. 
The most significant enhancement was observed in the D2A dataset~\cite{D2A}, with an increase of 7.26 points in the F1-score. However, in the Devign dataset~\cite{Devign}, the RAG module did not show an improvement. 
A closer examination of the data reveals that the datasets where the RAG module performed well generally consisted of shorter function codes, whereas the Devign dataset, where the module performed poorly, had much longer input sequences. 
This discrepancy is understandable, as longer function codes imply more complex problems and greater retrieval difficulty. 
The performance of the RAG module is therefore significantly influenced by the average length of the samples. 
Nonetheless, overall, the RAG module contributed to an average increase of 1.8 points of the F1-score. 
This indicates that the RAG module is effective, and it indeed solves, to some certain extent, the problem that the in-context learning setup cannot often obtain effective results on \textit{class-output} type tasks.


\find{{\bf \ding{45} Answer to RQ2: }$\blacktriangleright$ 
Traditional in-context learning often struggles with classification tasks. By incorporating a Retrieval-Augmented Generation (RAG) module, \tool significantly improves classification performance, achieving an average F1-score increase of 1.8 points compared to baseline in-context learning.
$\blacktriangleleft$ }


\begin{table*}[ht]
\centering
\caption{Performance comparison across different datasets.}
\label{table_fine}

\resizebox{.75\linewidth}{!}{
\begin{tabular}{cl|l>{\columncolor{black!20}}l|l>{\columncolor{black!20}}l|l>{\columncolor{black!20}}l|l>{\columncolor{black!20}}l}

\toprule

\multicolumn{1}{l}{} Task Type &Dataset& 7b    & \cellcolor{white}+ \tool & 7b-In & \cellcolor{white}+ \tool & 13b   & \cellcolor{white}+ \tool & 13b-In & \cellcolor{white}+ \tool \\ \midrule
\multirow{8}{*}{\begin{tabular}[c]{@{}c@{}}Code\\ Generation\end{tabular}}         & APPS~\cite{APPS}              & 3.27  & 17.1     & 13.9  & 23.55    & 4.41  & 15.65    & 3.91   & 27.39    \\
                                                                                   & CoNaLa~\cite{conala}            & 2.58  & 22.14    & 0.46  & 27.69    & 0.41  & 13.56    & 0.44   & 22.41    \\
                                                                                   & CONCODE~\cite{CONCODE}           & 1.1   & 24.82    & 0.95  & 12.33    & 0.69  & 12.2     & 1.65   & 13.35    \\
                                                                                   & MBPP~\cite{MBPP}              & 7.05  & 37.32    & 1.31  & 46.7     & 1.08  & 28.38    & 1.1    & 30.57    \\
                                                                                   & Mercury~\cite{Mercury}           & 0.67  & 43.3     & 0.05  & 28.52    & 3.34  & 33.06    & 0      & 45.1     \\
                                                                                   & InstructHumanEval~\cite{instructhumaneval} & 8.28  & 40.42    & 12.78 & 56.98    & 1.66  & 41.45    & 2.01   & 49.04    \\
                                                                                   & StudentEval~\cite{StudentEval}       & 4.95  & 45.91    & 1.32  & 57.07    & 1.1   & 49.8     & 1.13   & 58.01    \\
                                                                                   & XLCoST~\cite{XLCoST}            & 0     & 39.42    & 0     & 42.34    & 0     & 42.87    & 0      & 48.33    \\ \hline
\multirow{2}{*}{\begin{tabular}[c]{@{}c@{}}Code\\ Translation\end{tabular}}        & CodeXGLUE~\cite{CodeXGLUE}         & 1.38  & 31.64    & 1.34  & 85.23    & 1.46  & 85.73    & 1.52   & 87.42    \\
                                                                                   & HumanEvalX~\cite{HumanEval-X}        & 8.83  & 9.79     & 6.71  & 9.45     & 2.69  & 3.7      & 6.06   & 9.17     \\ \hline
\multirow{2}{*}{\begin{tabular}[c]{@{}c@{}}Code\\ Completion\end{tabular}}         & InstructHumanEval~\cite{instructhumaneval} & 11.3  & 46.93    & 16.81 & 38.81    & 1.23  & 49.23    & 3.29   & 41.39    \\
                                                                                   & SAFIM~\cite{safim}             & 1.55  & 52.47    & 0.45  & 67.08    & 0.4   & 76.08    & 0.71   & 75.86    \\ \hline
\multirow{2}{*}{\begin{tabular}[c]{@{}c@{}}Program\\ Repair\end{tabular}}          & CodeXGLUE~\cite{CodeXGLUE}         & 1.43  & 78.39    & 1.39  & 86.34    & 1.4   & 80.31    & 1.33   & 79.92    \\
                                                                                   & HumanEvalPack~\cite{HumanEvalPack}     & 4.27  & 97.16    & 3.59  & 95.36    & 2.05  & 96.67    & 2.56   & 97.03    \\ \hline
\multirow{3}{*}{\begin{tabular}[c]{@{}c@{}}Code\\ Summarization\end{tabular}}      & CodeXGLUE~\cite{CodeXGLUE}         & 0.44  & 2.23     & 0.43  & 18.27    & 0.48  & 22.06    & 0.48   & 28.93    \\
                                                                                   & XLCoST~\cite{XLCoST}            & 0     & 6.12     & 3.81  & 8.95     & 0     & 10.85    & 3.67   & 6.38     \\
                                                                                   & Funcom~\cite{Funcom}            & 0.14  & 23.87    & 0.15  & 22.35    &  0.15     &   18.1       &  0.17      &  18.79        \\ \hline
\begin{tabular}[c]{@{}c@{}}Clone\\ Detection\end{tabular}                          & BigCloneBench~\cite{BigCloneBench}     & 0     & 24.59    & 0     & 25.3     & 0     & 26.58    & 0      & 28.15    \\ \hline
\multirow{5}{*}{\begin{tabular}[c]{@{}c@{}}Vulnerability\\ Detection\end{tabular}} & Devign~\cite{Devign}            & 0 & 62.96    & 0 & 62.8     & 0 & 56.33    & 0  & 45.38    \\
                                                                                   & D2A~\cite{D2A}               & 0  & 68.15    & 0 & 68.23    & 0 & 55.46    & 0  & 54.78    \\
                                                                                   & BigVul~\cite{Big-Vul}            & 0     & 7.11     & 0     & 11.56    & 0     & 17.36    & 0      & 16.55    \\
                                                                                   & ReVeal~\cite{ReVeal}            & 0     & 12.05    & 0     &   19.73    & 0     &   18.56       & 0      &    19.92       \\
                                                                                   & Draper~\cite{Draper}            & 0      &  11.87        & 0      &     17.12     &  0     &   17.69 &  0   &   17.08       \\ \bottomrule
\end{tabular}}

\begin{tablenotes}
\footnotesize
    \item ``-In'' indicates the instruct version of the model.
    \item For code generation, code translation, code completion, program repair and code summarization tasks, the metric values in the table are BLEU.
    \item For clone detection and vulnerability detection tasks, the metric values in the table are F1-scores.
\end{tablenotes}
\vspace{-0.5cm}

\end{table*}

\subsection{RQ3: Performance on specific datasets}
\label{rq3}
This section examines \tool's performance across individual datasets to assess its effectiveness in various software engineering tasks. Results, summarized in Table~\ref{table_fine}, demonstrate \tool's ability to enhance model performance across all datasets.

The results reveal that \tool significantly enhances model performance across all datasets, albeit with varying degrees of effectiveness depending on the task type. For code-output tasks such as code generation, code translation, code completion, and program repair, the impact of \tool is most pronounced. For instance, in the APPS dataset for code generation, \tool boosts the performance of the 7b model from a BLEU score of 3.27 to 17.1, highlighting a substantial improvement. Similarly, in the CodeXGLUE dataset for program repair, the 7b model's performance leaps from 1.43 to 78.39 in terms of BLEU score.

In contrast, for the text-output task of code summarization, the improvements are less significant, though still notable. For example, in the CodeXGLUE dataset, the BLEU score for the 7b model increases from 0.44 to 2.23 with the application of \tool, indicating that while \tool provides some benefit, the task's inherent complexity or the model's baseline performance limits the magnitude of improvement.

For class-output tasks like clone detection and vulnerability detection, \tool proves essential as it enables models to perform tasks that they could not otherwise address effectively. In datasets like BigCloneBench and BigVul, non-\tool models fail to produce any valid outputs. Such models, however, when supported by \tool, achieve F1-scores of 24.59 and 7.11, respectively. This improvement underscores the importance of \tool in equipping models with the capability to handle classification tasks by providing relevant contextual information via the RAG module.

Overall, \tool's impact varies across tasks and datasets, influenced by factors such as dataset complexity, input length and the intrinsic nature of the tasks.

\find{{\bf \ding{45} Answer to RQ3: }$\blacktriangleright$ 
\tool significantly improves model performance across a range of software engineering tasks. The most significant gains are observed in code generation and program repair, whereas improvements in code summarization are more modest. A key value of \tool lies in enabling models to effectively tackle classification problems where traditional in-context learning falls short.$\blacktriangleleft$ }
\subsection{RQ4: Case study}
\label{rq4}

\begin{figure*}
    \centering
    \includegraphics[width=\linewidth]{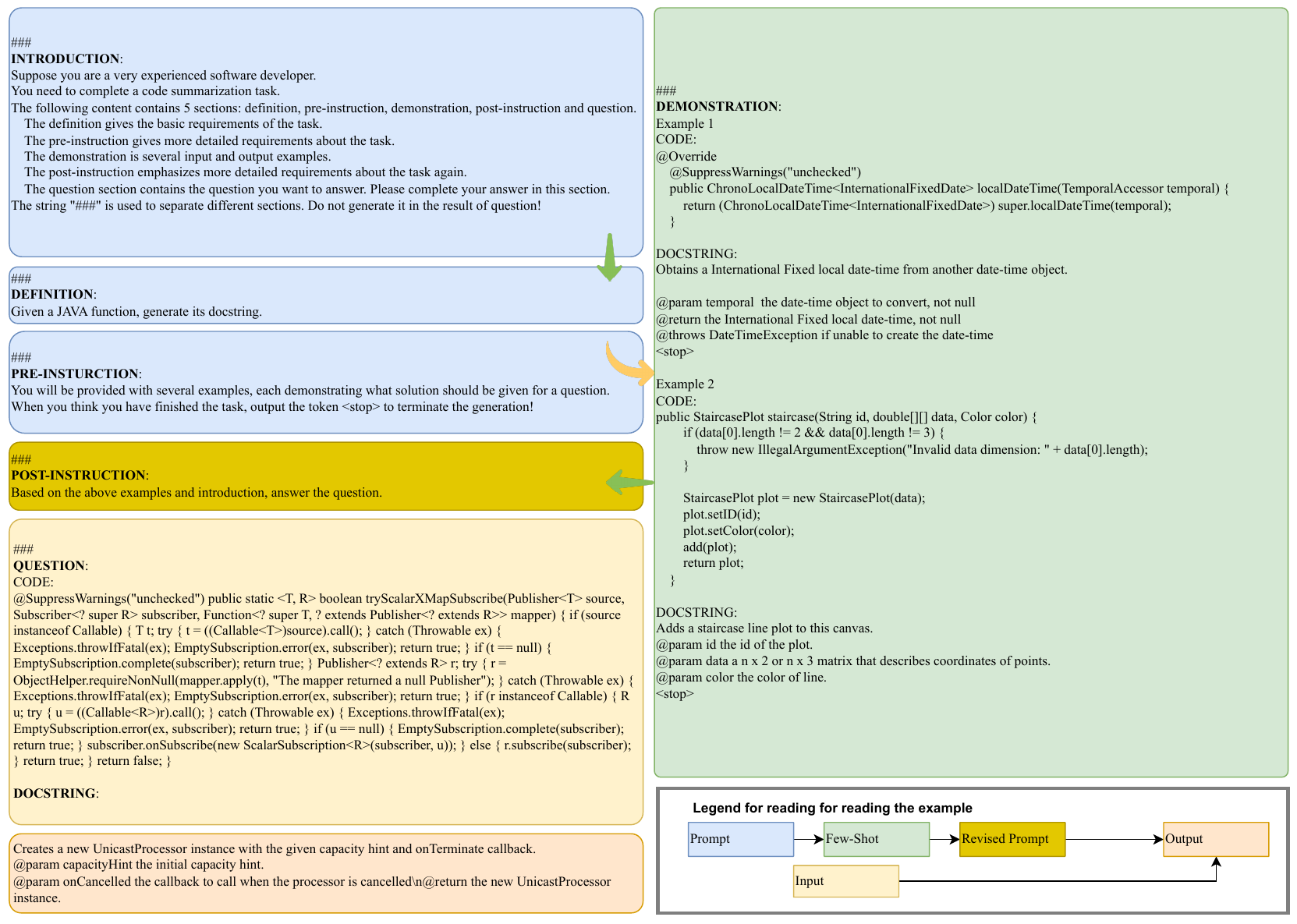}
    \caption{Illustration of \tool's prompt and output on the code summarization task.}
    \label{fig:casetext}
\end{figure*}

We provide a specific example to illustrate and analyze the Prompt constructed by \tool and the content of it outputs. Taking the code summarization task as an example. It is a text-output task where, given a piece of code, the task consists in providing a description of the code in natural language. We can see the specific prompt content and output content of \tool in Figure~\ref{fig:casetext}. 
First, the prompt contains the six sections mentioned in Sec.\ref{configurable_prompt}. 
In the {INTRODUCTION} section, we provide the scenario, set the role of an experienced software developer for the LLM, and briefly introduce the sections that will appear in the entire prompt. 
Then, in the {DEFINITION}, we give the basic requirements of the task, which is to generate a docstring for a given JAVA function. 
Next, in the {PRE-INSTRUCTION}, we emphasize some content that the model needs to pay attention to. Then comes the {DEMONSTRATION} section, which is the longest and most important section of the entire prompt. 
It provides the model with several input and output examples (taken from the training set), allowing the model to directly learn the ability to handle the target task within the prompt. 
After that, in the {POST-INSTRUCTION} section, we emphasize again the issues that the model needs to pay special attention to. 
Finally, the {QUESTION} section is the input part of a certain test data in the test set, and we hope that the model can answer this question. It can be seen that the model correctly outputs the docstring of the function given in the {QUESTION}.

Through this example, we can see that the prompt constructed by \tool provides the model with rich in-context knowledge, including situation setting ({INSTRUCTION}), task definition ({DEFINITION}), cautions ({PRE-INSTRUCTION} and {POST-INSTRUCTION}), and question-answering examples ({DEMONSTRATION}). Among these, the {DEMONSTRATION} part usually occupies the largest proportion of the prompt. These contents provide sufficient reference for the model to answer new questions and meticulously constrain the model's output. Therefore, it is not difficult to understand why the model can still exhibit a certain ability to solve problems even when it is not configured according to the prompt template of the instruct version model itself. Furthermore, \tool's default prompt construction scheme can not only adapt to the task given in the example, but the previous experimental results show that it can solve various code-text problems, making it highly generalizable.

\find{{\bf \ding{45} Answer to RQ4: }$\blacktriangleright$ 
The case study on the code summarization task highlights how \tool's structured prompt design provides rich in-context learning opportunities.
$\blacktriangleleft$ }

\section{Discussion}


\subsection{Threats to Validity}

The internal threat to validity lies in the implementation of the framework and the result of experiments. First, to ensure the robustness of the implementation, our framework's inference feature and RAG feature are powered by widely recognized projects \textbf{VLLM}~\cite{vllm} and \textbf{Llama-index}~\cite{llamaindex}, respectively. This avoids the uncontrollable factors introduced by repeated implementation. Then, in terms of experiments, we conducted a large number of tests with different configurations, including replacing the text in each section of the prompt, changing the number of shots in the demonstration, adjusting the sampling rate of the RAG module for obtaining candidates from the training set, etc., in order to achieve the best model performance.

The external threat to validity lies in the correctness of the parameters and data of the models we used. To address this issue, all of our models (\textbf{CodeLLAMA} series used for the generator and \textbf{CodeBERT} for the RAG module) and datasets are downloaded from the officially open-sourced repositories on the Huggingface platform.

\subsection{Limitations}

The limitations of our proposed method are mainly reflected in the diversification of the base model and the efficiency of the RAG module. Due to the limitations of experimental resources, we only conducted experiments using the 7b and 13b models of \textbf{CodeLLAMA} series as the base models. However, there are many state-of-the-art SE-related LLMsavailable to use as base models to make a stronger case for \tool. Moreover, although our framework can enable the model to handle code-text related problems without any fine-tuning (or training), which saves a lot of time, for tasks that require the RAG module to be completed effectively, our method still incurs a substantial time overhead.

\subsection{Why not fine-tuning in \tool}
In this work, we opted not to fine-tune CodeLLAMA for several reasons. First, fine-tuning is typically tailored to specific tasks, which limits the model's ability to generalize across different domains or adapt to new tasks without retraining. This specificity can reduce the versatility of the model, making it less suitable for scenarios requiring adaptability to various software engineering tasks. Second, the computational cost and resource demands of fine-tuning large models like CodeLLAMA are significant, often necessitating specialized hardware and extended training times. This can be prohibitive in terms of both time and financial resources, especially when compared to leveraging in-context learning, which allows the model to utilize its pre-trained knowledge and adapt on-the-fly with task-specific prompts. By avoiding fine-tuning, we maintain the flexibility and efficiency of our approach, allowing for rapid adaptation to a wide range of tasks without the overhead associated with fine-tuning processes.

\section{Related Work}
\label{sec:related_work}


\textbf{Code LLMs} such as CodeT5~\cite{wang2021codet5}, Code-Gen~\cite{nijkamp2022codegen}, CODELLAM~\cite{roziere2023code}, and DeepSeek-Coder~\cite{guo2024deepseek}, are pre-trained over a largecodebases from scratch, and show some abilities for general code generation and understanding. 
There have since been extensive studies fine-tuning task-specific Code LLMs to perform software engineering tasks like automatic program repair~\cite{xia2023conversational, xia2023automated}, code translation~\cite{pan2023understanding, chen2024data}, and code summarization~\cite{wang2023codet5+}. 
Furthermore, these base code models are also finetuned~\cite{luo2023wizardcoder, zhang2023repocoder} or
prompted~\cite{sun2023prompt, zhou2022large} to unlock their full potential by specializing them in solving domain-specific coding tasks. While such code-text pre-training has become mainstream, in-context learning for this bimodality remains under-explored. This paper introduces a comprehensive study of code-text in-context learning, focusing on leveraging pre-trained large language models.

\textbf{Few-shot prompting} is now a popular technique for enhancing LLM performance on specific tasks. By providing a few examples within the prompt, models can adapt to new tasks without extensive training. While previous research has successfully applied few-shot prompting to various software engineering downstream tasks~\cite{geng2024large, ahmed2024automatic, feng2024prompting, first2023baldur, kang2023large, wei2022clear}, our work introduces a novel approach: a general-purpose interface that eliminates the need for task-specific training. Unlike traditional few-shot prompting, which relies on carefully curated examples, our method offers a more flexible and adaptable solution for handling diverse SE tasks.

\section{Conclusion}
We introduced \tool, a novel framework that addresses the challenges inherent in building general-purpose code-text models. By leveraging in-context learning and a unified code-text interface (i.e., adapter), \tool empowers LLMs to effectively handle a wide range of software engineering tasks without requiring extensive retraining. Our empirical evaluation with the CodeLLAMA base models demonstrates \tool's ability to significantly enhance model performance across diverse tasks, particularly in code generation and program repair. A key component, the Retrieval-Augmented Generation module, further enables \tool to support CodeLLAMA on some software engineering classification tasks where it was ineffective.

While this research represents a step forward in bimodal software engineering, there are several avenues for future exploration. These include investigating the cost of InCTRL vs the cost of task-specific fine-tuning, and developing methods for evaluating model robustness.

\noindent{\bf Open Science:} All artefacts are available at:
\begin{center}
        {\url{https://anonymous.4open.science/r/inctrl-B65B}}
\end{center}

\newpage

\bibliographystyle{IEEEtran}
\balance
\bibliography{references, datasets, refref, icl}

\end{document}